\documentclass[useAMS,usenatbib]{mn2e}

\usepackage{graphics}
\usepackage{epsfig}
\usepackage{color}
\usepackage{xspace}

\title[Decay of the vacuum energy into CMB photons]{Decay of the vacuum energy into CMB photons}
\author[Reuven Opher and Ana Pelinson]{Reuven Opher\thanks{E-mail:
opher@astro.iag.usp.br} and Ana Pelinson\thanks{E-mail:
anapel@astro.iag.usp.br}\\
IAG, Universidade de
S\~{a}o Paulo, Rua do Mat\~{a}o 1226, \\
Cidade Universit\'aria, CEP 05508-900 S\~{a}o Paulo, SP, Brazil }
\begin{document}

\date{2005 June 3}
\maketitle

\begin{abstract}
We examine the possibility of the decay of the vacuum energy into a
homogeneous distribution of a thermalized cosmic microwave
background (CMB), which is characteristic of an adiabatic vacuum
energy decay into photons. It is shown that observations of the
primordial density fluctuation spectrum, obtained from CMB and
galaxy distribution data, restrict the possible decay rate. When
photon creation due to an adiabatic vacuum energy decay takes place,
the standard linear temperature dependence $T(z)=T_0\,(1+z)$ is
modified, where $T_0$ is the present CMB temperature, and can be
parameterized by a modified CMB temperature dependence
$\overline{T}(z)=T_0\,(1+z)^{1-\beta}$. From the observed CMB and
galaxy distribution data, a strong limit on the ma\-ximum value of
the decay rate is obtained by placing a maximum value $\beta_{{\rm
max}}\simeq 3.4 \times 10^{-3}$ on the $\beta$ parameter.
\end{abstract}

\section{Introduction}

The present observed acceleration of the universe is due to a
substance which we call dark energy, the nature of which is yet
unknown. \citet{brons} was the first to introduce the idea of the
possible decay of dark energy. A recent review of possible
explanations for the nature of dark energy and its possible decay
can be found, for example, in \citet{vdec7c}. As noted by
\citet{vdec7c}, the evolution of the dark energy density and its
related coupling to matter or radiation is, in general, assigned and
not derived from an action principle. Discussions of this can be
found in \citet{pol,kaz,vdec1b,gas,sato,bart,vdec4a,maty,birk}.

In the present paper, we assume that the dark energy is the vacuum
energy and investigate its possible decay. Some scalar field dark
energy models, such as that of \citet{pb1} and \citet{pb2}, were
motivated, in part, by particle physics and used observational data
to constrain the decay rate of dark energy. However, in general,
almost all studies of the decay of the dark energy, assuming that it
is the vacuum energy, as is done in this paper, are purely
phenomenological and do not put strong limits on the decay from
observational data (see, for example,
\citealt{vdec0,vdec0a,vdec0c,vdec1,vdec1a,vdec1b,vdec1d,vdec2,vdec2b,vdec2c,vdec3a,
vdec4,vdec4a,vdec5,vdec5a,vdec5b,vdec5c,vdec5d,vdec6,vdec7,vdec7a,vdec7b,vdec7c,vdecaycoment}).

In a previous paper, we studied the limit put on the rate of a
possible vacuum energy decay into cold dark matter (CDM) by the
observed cosmic microwave background (CMB) and large galaxy survey
data \citep{vdecay}. The observed temperature fluctuations of the
CMB photons $(\delta T/T)^2$ are approximately proportional to CDM
density fluctuations $(\delta \rho/\rho)^2$ \citep{padm}. CDM
density fluctuations derived from the CMB data can be compared with
those derived from the 2dF Galaxy Redshift Survey (2dFGRS)
\citep{2df,2df1}. It was found that the $(\delta \rho/\rho)^2$
derived from the galaxy distribution data differs from the $(\delta
\rho/\rho)^2$ derived from the CMB data by no more than 10 per cent
\citep[see][for the final data set of the 2dFGRS]{2dfend}.

A vacuum energy decaying into CDM increases its total density,
diluting $(\delta\rho/\rho)^2$. In order to evaluate $(\delta
\rho/\rho)^2$ at the recombination era, when it created the $\delta
T/T$ of the CMB, its present measured value obtained from the galaxy
distribution data must then be increased by a factor $F$. Since the
$(\delta \rho/\rho)^2$ derived from the CMB and galaxy distribution
data agree to 10 per cent, the maximum value for $F$ is $F_{\rm
{max}}=1.1$.

We found that the decay of the vacuum energy into CDM as a scale
factor power law $\rho_{\Lambda}\propto (1+z)^{n}$, gives a maximum
value for the exponent $n_{\rm max}\approx 0.06$. For a parametrized
vacuum decay into a CDM model with the form
$\rho_{\Lambda}(z,\nu)=\rho_{\Lambda}(z=0)+\rho_{\rm{c}}^0\,[\nu/(1-\nu)]\,
[\left(1+z\right)^{3(1-\nu)}-1]\, \,$, where $\rho_{\rm{c}}^0$ is
the present critical density, an upper limit on the $\nu$ parameter
was found to be $\nu_{\rm max}=2.3\times 10^{-3}$.

Here, we study the limit imposed on the rate of a possible decay of
the vacuum energy into a homogeneous distribution of thermalized CMB
photons. In this scenario, $(\delta T/T)^2$ at the recombination
epoch were diluted by photons created by a vacuum energy decay.
Thus, $(\delta T/T)^2$ at present are smaller than those existing at
the recombination era. This implies larger $(\delta\rho/ \rho)^2$ at
the recombination era than those derived from the observed CMB data.

We generally assume that the CMB temperature $T$ is proportional to
$(1+z)$ in a Friedmann-Robertson Walker (FRW) universe. There is,
however, little direct observational evidence for this relation
despite considerable observational efforts to verify it.
\citet{Lima1} summarized some of the observational studies which
have been made up to a redshift $z\sim 4.5$.

Although in the present paper, we investigate a possible decay of a
homogeneous vacuum energy into a homogeneous distribution of
photons, the vacuum energy may, in principle, not be homogeneous and
its decay could then lead to an inhomogeneous distribution of
photons. A vacuum energy depending on spatial position could be
created, for example, by a Casimir effect, such as that described by
Muller, Fagundes and Opher (2001,2002). The inhomogeneous
distribution of decay photons produced might be able to be detected
in high precision CMB data. This possibility will be investigated in
a future study.

In Section II, we put strong limits on the possible decay of the
vacuum energy into a homogeneous distribution of thermalized CMB
photons and its effect on $(\delta \rho/\rho)^2$ derived from the
observed CMB data. Our conclusions are presented in Section III.


\section{Vacuum energy decay into CMB photons}

According to the standard model, $(\delta T/T)^2$ were created at
${z}_{\rm rec}\sim 1100$, the recombination epoch \citep{padm}. In
the standard model,  $(\delta T/T)$ observed today are given by the
expression
\begin{equation}
\left(\frac{\delta T}{T}\right)\,\Big|_{z\sim {0}}\,= {\cal
K}\,\,\,\frac{\delta\rho}{\rho}\,\Big|_{{{z}_{\rm{rec}}}}\,,
\label{zto1070}
\end{equation}
where ${\cal K}$ is approximately constant and the temperature
dependence of $T(z)$ is
\begin{equation}
{T}(z)=T_0\left(1+z\right)\,, \label{tempstan2}
\end{equation}
where $T_0\simeq 2.75\,\rm{K}$ is the present CMB temperature. The
present value of $(\delta \rho/\rho)^2$ is gotten  from the relation
\begin{equation}
\left(\frac{\delta\rho}{\rho}\right)\,\Big|_{z\sim {0}}\,={\cal
D}\,({z}_{\rm rec}\rightarrow z=0)
\,\,\,\frac{\delta\rho}{\rho}\,\Big|_{{z}_{\rm rec}}\,,
\label{phizto1070}
\end{equation}
where ${\cal D}\,({{z}_{\rm rec}}\rightarrow z=0)$ is the growth
factor from the recombination era until the present time.

Let us examine a possible vacuum decay into photons. Assuming that
the decay is adiabatic, the vacuum ener\-gy decays into a
homogeneous distribution of thermalized black body CMB photons. This
was shown by \citet{Lima1}, as follows. The conservation equation
for the photon number density $n$ is
\[
\dot{n}+3nH=\psi\,, \] where $H$ is the Hubble parameter and $\psi$
is the photon source term. Taking Gibbs law into account, we have
\[
n\,T\,\rm{d}\sigma =\rm{d}\rho -\frac{\rho +P}{n}\,\rm{d}n\,, \]
where $\sigma$ is the specific entropy. Since $\rm{d}\sigma$ is an
exact differential, we obtain the thermodynamic identity
\[
T\,\left(\frac {\partial P}{\partial T}\right)_{n}=\rho
+P-n\left(\frac {\partial \rho }{\partial n}\right)_{T}\,.
\]
Using $T$ and $n$ as independent thermodynamic variables, we obtain
\[
\frac{\dot{T}}{T}=\left( \frac{\partial P}{\partial T}\right) _{n}%
\,\frac{\dot{n}}{n}-\frac{\psi }{n\,T\left({\partial \rho }/{%
\partial T}\right)_n }\left[ \rho +P-\frac{n\,\dot{\rho }_{V}}{\psi }%
\right]\,,
\]
\[
\dot{\sigma }=\frac{\psi }{n\,T\left( \frac{\partial \rho }{\partial T}%
\right) _{n}}\left[ \rho +P-\frac{n\,\dot{\rho }_{V}}{\psi
}\right]\,,
\]
where $\dot{\rho }_{V}$ is the decay rate of the vacuum energy
density \citep{vdec5d}. For photons, we have $P=\rho/3$. In order to
have homogeneous equilibrium black body radiation, $n\propto T^3$,
we must have
\[
\rho +P-\frac{n\,\dot{\rho }_{V}}{\psi }=0
\]
or
\[ \dot{\rho }_{V}=\frac{4\rho }{3n}\psi\,.
\]
We then have
\[ \dot{\sigma}=0\,.
\]
Thus, when the decay of the vacuum energy is adiabatic and the
specific entropy does not change, we obtain a homogeneous thermal
distribution of black-body radiation. From the conservation equation
for the photon density, $\dot{n}+3nH=\psi\,$, and the above relation
between $\dot{\rho }_{V}$ and $\psi$, we have
\[
\frac{\dot{T}}{T}=-H+\frac{\psi }{3n}\,.
\]
For $\psi=0$ (no photon creation), we obtain the standard FRW law
relation, [Eq.(\ref{tempstan2})].

It is to be noted that previous studies of the decay of the vacuum
energy assumed that the decay rate into photons $\psi$ is
proportional to some power of $H$ and/or the cosmic scale factor $a$
(i.e., $\psi\propto H^{\alpha}\,a^{\gamma}$, where $\alpha$ and
$\gamma$ are constants). The combination of the values $\alpha=1$
and $\gamma=-3$, is especially interesting as it indicates an
adiabatic decay of the vacuum energy into a homogeneous distribution
of thermalized CMB photons. Since $n$ is inversely proportional to
$a^{-3}$, we have $\psi \propto H\,n$. Defining
$\beta=\psi/3\,n\,H$, we obtain
\begin{equation}
\overline{T}(z)=T_0\left(1+z\right)^{1-\beta}\, . \label{tempbeta}
\end{equation}
According to \cite{Lima1}, the possible range of $\beta$ is $\beta
\in [0,1]$. The aim of the present article is to put a strong upper
limit on the possible value of $\beta$.

Two effects are produced by the decay of the vacuum energy into CMB
photons:
\begin{description}
\item [1)]There is a decrease in the observed $(\delta
T/T)^2$ due to the increase of the homogeneous distribution of black
body photons from the vacuum decay; and
\end{description}

\begin{description}
\item [2)]When there is a vacuum energy decay into CMB photons, the
value of the recombination redshift ${\bar{z}_{\rm rec}}$ is higher
than that of the standard model ${{z}_{\rm rec}}$ since the universe
is cooler at any given redshift. The recombination temperature thus
occurs at a higher $z$.
\end{description}

Due to the dilution of $( {\delta T}/{T}),$ instead of
Eq.(\ref{zto1070}) of the standard model, we must use the relation
\begin{equation}
F_1\,\left(\frac{\delta T}{T}\right)\,\Big|_{{{z}_{\rm rec}}}\,=
{\cal K}\,\,\frac{\delta\rho}{\rho}\,\Big|_{{{z}_{\rm
rec}}}\,.\label{first}
\end{equation}
We define
\begin{equation}
F_{1}(z)\equiv \left[ \frac{{T}\,(z)}{{T}%
\,(z)-\Delta T(z)}\right] \,\Big|_{{{z}_{\rm rec}}}\, \,.
\label{factort}
\end{equation}
The difference between the recombination temperature ${T} ({z}_{\rm
rec})$ predicted by the standard model and that of the model in
which the vacuum energy decays into photons at temperature
$\overline{T} ({z}_{\rm rec})$ is
\begin{equation}
\Delta T({{z}_{\rm rec}})={T}\,({{z}_{\rm rec}})-\overline{T}
({z}_{\rm rec})\,. \label{deltat2}
\end{equation}
Using Eqs.(\ref{tempbeta}), (\ref{factort}), and (\ref{deltat2}), we
have
\begin{equation}
F_1=\left(1+{{z}_{\rm rec}}\right)^{\beta}\label{f1beta}\,.
\end{equation}

From Eqs (\ref{tempbeta}) and (\ref{deltat2}), $\overline{T} (z)$
was lower than ${T}\,(z)$ by $\Delta T$ at ${z}_{\rm rec}$. Thus,
the resultant recombination redshift $\bar{z}_{\rm rec}$ was higher
than that of the standard model ${{z}_{\rm rec}}$. Instead of
Eq.(\ref{phizto1070}), $\left({\delta\rho}/{\rho}\right)\,$ at
${z\sim {0}}\,$ is now given by
\begin{equation}
\left(\frac{\delta\rho}{\rho}\right)\,\Big|_{z\sim {0}}\,={\cal
D}\,(\bar{z}_{\rm rec}\rightarrow z=0)
\,\,\,\frac{\delta\rho}{\rho}\,\Big|_{z=\bar{z}_{\rm rec}}\,,
\label{phiztorec}
\end{equation}
where ${\cal D}\,(\bar{z}_{\rm rec}\rightarrow z=0)$ is the density
fluctuation growth factor from the recombination era at
$\bar{z}_{\rm rec}$ until the present epoch, due to the decay of the
vacuum energy into photons.  Therefore, instead of
Eq.(\ref{zto1070}), we have
\begin{equation}
\left(\frac{\delta T}{T}\right)\,\Big|_{z\sim {0}}\,= {\cal
K}\,\,\,\frac{\delta\rho}{\rho}\,\Big|_{{z}=\bar{z}_{\rm rec}}\,.
\label{ztorec}
\end{equation}

Using Eqs.(\ref{phizto1070}) and (\ref{first}), we have
\begin{equation}
\left(\frac{\delta\rho}{\rho}\right)\Big|_{z\sim
{0}}=\frac{F_1}{{\cal K}}\, {\cal D}({{z}_{\rm rec}}\rightarrow
z=0)\left(\frac{\delta T}{T}\right)\Big|_{{{z}_{\rm rec}}}
\label{transf1}
\end{equation}
and from Eqs.(\ref{phiztorec}) and (\ref{ztorec}),
\begin{equation}
\left(\frac{\delta\rho}{\rho}\right)\,\Big|_{z\sim
{0}}\,=\frac{F_1}{{\cal K}}\, {\cal D}(\bar{z}_{\rm rec}\rightarrow
z=0)\left(\frac{\delta T}{T}\right)\Big|_{{{z}_{\rm rec}}} \,.
\label{transf2}
\end{equation}

Equations (\ref{transf1}) and (\ref{transf2}) give the correction
factor $F_2$ due to the change in the value of the recombination
redshift,
\begin{equation}
F_2=\frac{{\cal D}(\bar{z}_{\rm rec}\rightarrow z=0)}{{\cal
D}({{z}_{\rm rec}}\rightarrow z=0)}\,.\label{f2}
\end{equation}
The growth of a perturbation in a matter-dominated Einstein-de
Sitter universe is $\delta\rho/\rho\propto a = (1+z)^{-1}$, where
$a$ is the cosmic scale factor \citep[see e.g., ][]{coles}. Thus,
the growth factor $\cal D$ is
\[
{\cal D}\simeq (1+z)\,.
\]
We then find from Eq.(\ref{f2})
\begin{equation}
F_2\simeq \left(\frac{1+\bar{z}_{\rm rec}}{1+{z}_{\rm
rec}}\right)\,. \label{f2z}
\end{equation}

The temperature at ${{z}_{\rm rec}}$ in the standard model is
\begin{equation}
{T}\,({{z}_{\rm rec}})=T_0\,(1+{{z}_{\rm rec}})\,.
\end{equation}
In order for the temperature at the recombination epoch
$\bar{z}_{\rm rec}$, when the vacuum energy is decaying into CMB
photons,  to be $T ({z}_{\rm rec})$, we must have, from
Eq.(\ref{tempbeta}),
\begin{equation}
\bar{z}_{\rm rec}=(1+{{z}_{\rm rec}})^{1/(1-\beta)}-1\,.
\label{zdecay}
\end{equation}
From Eq.(\ref{f2z}), we then have
\begin{equation}
F_2\simeq (1+{{z}_{\rm rec}})^{{\beta}/{(1-\beta)}}\,.
\label{f2comp}
\end{equation}

The total factor $F$ is composed of $F_1$, due to the dilution of
the CMB as a result of vacuum energy decay, and $F_2$, due to the
change in the redshift of the recombination epoch. Assuming that the
effects described by $F^2_1$ and $F^2_2$ are independent and that
the total factor $F$ is the product of $F_1^2$ and $F_2^2$, we have
\begin{equation}
F=F_1^2\,F_2^2\,. \label{factordecay}
\end{equation}

Thus, from Eqs.(\ref{f1beta}), (\ref{f2comp}) and
(\ref{factordecay}), the condition for the maximum value of $\beta
\in [0,1]$ is
\begin{equation}
{\beta}_{\rm max}= \alpha \left[ 1- \sqrt{ 1- \frac{ \ln{(F_{\rm
max})} } {2{\alpha}^2 \ln{(1+ {z}_{\rm rec}) } }    }\, \right]\,,
\label{betaeq}
\end{equation}
where
\begin{equation}
\alpha=1+\frac{ \ln{(F_{\rm max})} } {4 \ln{(1+ {z}_{\rm rec}) }
}\,. \label{alpha}
\end{equation}

\begin{figure*}
\includegraphics[scale=1.0]{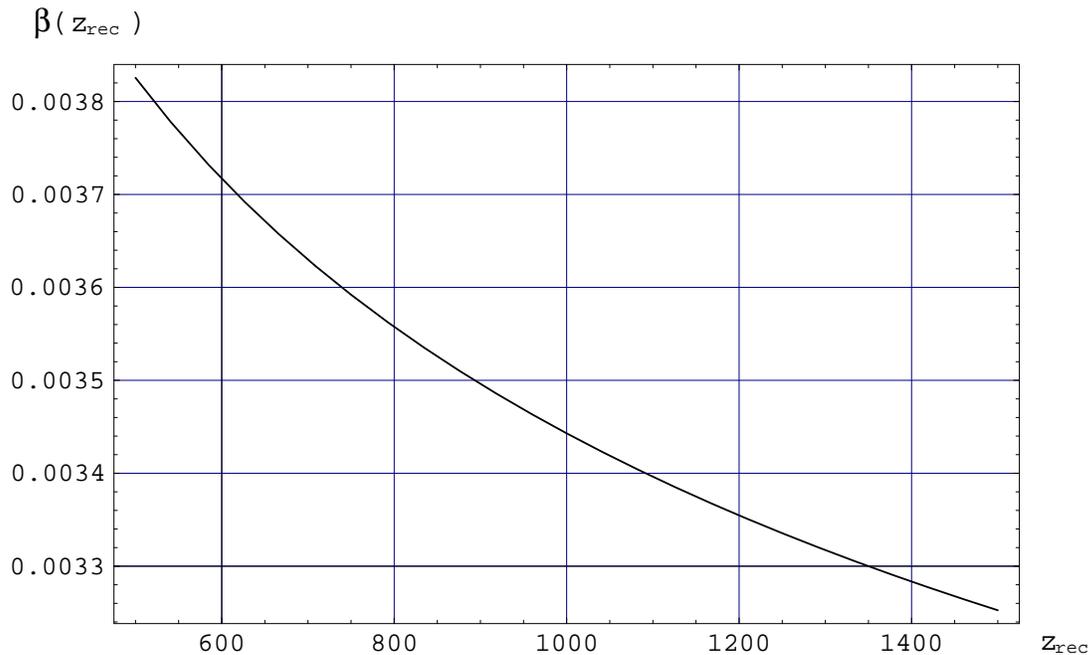}
\caption{The dependence of the $\beta$ parameter on ${z}_{\rm rec}$
for $F=F_{\rm max}=1.1$ from Eqs.(\ref{betaeq}) and (\ref{alpha}). }
\label{zrecvar}
\end{figure*}

As noted above, the maximum value of $F$ from observations is
$F_{\rm max}=1.1$. A plot of $\beta$ vs ${z}_{\rm rec}$ in the
standard model is shown in Fig.~\ref{zrecvar} for $F=F_{\rm
max}=1.1$ from Eqs.(\ref{betaeq}) and (\ref{alpha}). For ${z}_{\rm
rec}\simeq 1100$, we find a very small maximum value of the $\beta$
parameter, $\beta_{\rm max}\simeq 3.4 \times 10^{-3}\,.$

\section{Conclusions}

We show that the CMB data, together with the large galaxy survey
data, put strong limits on the rate of a possible decay of the
vacuum energy into a homogeneous distribution of thermalized CMB
photons, between the recombination era and the present. Using the
fact that the $(\delta \rho/\rho)^2$ derived from the CMB and galaxy
distribution data do not differ by more than 10 per cent, we can
place an upper limit on the $\beta$ parameter for the decay of the
vacuum energy into CMB photons, parametrized by a change in the CMB
temperature at a given redshift $z$:
$\,\overline{T}(z)=T_0(1+z)^{1-\beta}\, $. We find that $\beta_{\rm
max}\simeq 3.4 \times 10^{-3}$.

In the above analysis, we assumed an Einstein-de Sitter CDM universe
with a growth factor ${\cal D}(z)\simeq 1+z$ to obtain $\beta_{\rm
max}$. This is true in a pressureless universe with an equation of
state $P/\rho\equiv w\simeq 0$, where $P$ is the pressure and $\rho$
is the energy density. ${\cal D}(z)$ will change at small redshifts
if $w(z)$ becomes less than zero due to a vacuum energy contribution
$(w_{{\rm V}}=-1)$, a quintessence energy contribution $(-1<w_{{\rm
Q}}<0)$, or a phantom energy contribution $(w_{{\rm P}}< -1)$. As a
result, $\beta_{\rm max}$ will change somewhat. An investigation of
these contributions is left for a future article.

Our results indicate that the rate of the decay of the vacuum energy
into CMB photons is extremely small. They are consistent with a zero
vacuum energy decay, $\beta=0$.

\section*{Acknowledgments}

R.O. thanks the Brazilian agencies FAPESP (00/06770-2) and CNPq
(300414/82-0) for partial support. A. P. thanks FAPESP for financial
support (03/04516-0 and 00/06770-2). The authors would like to thank
J.M. Overduin for his comment on a preliminary version of the
manuscript.

\end{document}